\documentclass[useAMS]{mn2e}
\usepackage{lscape}
\usepackage{rotating}

\def\cm3{\hbox{cm$^{-3}$}}
\def\lm{\hbox{$L_{\rm V}/M_{\rm {dyn}}$}}

\input psfig.sty
\input epsf.sty
\usepackage{graphicx}
\title[A Detailed Study of the Enigmatic Cluster M82F]
{A detailed study of the enigmatic cluster M82F}
\author[Bastian et al.] {N. Bastian$^1$, I. Konstantopoulos$^1$, L.J. Smith$^{1,2}$, G. Trancho$^{3,4}$, \newauthor M. S. Westmoquette$^1$, J.S. Gallagher III$^5$ \\
$^1$ Department of Physics and
Astronomy, University College London, Gower Street, London, WC1E 6BT\\
$^2$ Space Telescope Science Institute and European Space Agency, 3700 San        
Martin Drive, Baltimore, MD 21218, USA \\            
$^3$ Gemini Observatory, 670 N. A'ohoku Place, Hilo, HI 96720, USA\\
$^4$ Universidad de La Laguna, Tenerife, Canary Islands, Spain \\
$^5$ Department of Astronomy, University of Wisconsin-Madison, 5534 Sterling, 475 North Charter Street, Madison, WI 53706, USA\\
}
\date{Accepted. Received; in original form}
\pagerange{\pageref{firstpage}--\pageref{lastpage}}
\pubyear{2006}
\begin{document}
\maketitle
\label{firstpage}
\begin{abstract}
We present a detailed study of the stellar cluster M82F, using
multi-band high resolution HST imaging and deep ground based optical
slit and integral field spectroscopy.  Using the imaging we create colour maps of the cluster
and surrounding region in order to search for substructure.  We find a
large amount of substructure, which we
interpret as the result of differential extinction across the projected face of the
cluster.  With this interpretation, we are able to construct a spatially resolved extinction
map across the cluster which is used to derive the intrinsic flux distribution.  Fitting cluster profiles (King
and EFF) to the intrinsic images we find that the cluster is 15-30\%
larger than previous estimates, and that no strong evidence of mass
segregation in this cluster exists.  Using the optical spectra, we
find that the age of M82F is $60-80$~Myr and from its velocity
conclude that the cluster is not physically associated with a large
H~{\sc ii} region that it is projected upon, both in agreement with previous studies.  The reconstructed integral field maps show that that majority of the line emission comes from a nearby H~{\sc ii} region.  The spatial dependence of the line widths (implying the presence of multiple components) measured corresponds to the extinction map derived from photometry, indicating that the gas/dust clouds responsible for the extinction are also partially ionised. Even with the wealth of observations presented here, we do not find a conclusive solution to the problem of the high light-to-mass ratio previously found for this cluster and its possible top-heavy stellar IMF.

\end{abstract}
\section{Introduction}\label{intro}

The young massive cluster M82F in the starburst galaxy M82 has
received a considerable amount of attention in recent years.  It has
been the focus of medium and high resolution, optical and near-IR
spectroscopy as well as high-resolution, space-based imaging
(Gallagher \& Smith~1999, Smith \& Gallagher~2001, McCrady, Graham, \&
Vacca~2005, Bastian \& Goodwin~2006).  These observations have produced some surprising results.  These results can be summarised as follows.  The age of M82F is $60\pm20$~Myr despite being spatially coincident with an H~{\sc ii} region.  The stellar IMF within the cluster is extremely top-heavy with many fewer  stars below $2-3~M_{\odot}$  than expected for a normal IMF, while the radius appears to change as a function of wavelength, possibly indicating a large amount of mass-segregation.   Additionally, M82F displays an excess of light at large radii indicating that the cluster may not be in dynamical equilibrium.

Smith \& Gallagher~(2001, hereafter SG01) found that although M82F is
spatially coincident with a large H~{\sc ii} region (in projection) the
cluster velocity seems to differ from that of the H~{\sc ii} region by $\sim25$~km/s, as inferred from emission lines, and
hence may not be physically associated with it.  These authors (also see
Gallagher \& Smith 1999, hereafter GS99) conclude that M82F is deep within the galaxy
and that we are seeing the cluster through a chance line of sight of relatively low
extinction.  The magnitude of the extinction is also a
matter of some debate, as different authors obtain discrepant
values of the amount of extinction for M82F, with SG01 and McCrady et
al.~(2005, hereafter MGV05) reporting A$_{V}$ values of 2.8 mag and
$0.6-1.1$ respectively. 

The estimate of extinction is one of the essential ingredients when attempting
to calculate the stellar IMF by comparing the ratio of 
light to dynamical mass estimates, with that of simple stellar
population models (SSPs, all stars are coeval and assumed to have the same metallicity).  Other required  parameters are the age,
velocity dispersion, effective radius, distance to the source, and the
observed magnitude.  Once these ingredients are assembled, and
assuming that the cluster is in virial equilibrium, one can
place the cluster in the luminosity to dynamical mass ($L/M_{dyn}$) vs
age diagram along with  the corresponding predictions of simple stellar population models
composed of varying IMFs.

Using the $L/M_{dyn}$ as a diagnostic, SG01
and MGV05 have concluded that M82F is unique in that it may
have an extremely top-heavy stellar IMF  as it contains many fewer stars below 2 or
3~$M_{\odot}$ than predicted by a standard Salpeter (1955) IMF.  Other explanations of the derived position in the $L/M_{dyn}$
vs age diagram are errors in the age, radius, velocity dispersion, or
extinction in previous studies or that the cluster is not in virial
equilibrium.   Goodwin \& Bastian~(2006) have shown that M82F is also
unique in the age vs. $L/M_{dyn}$ plane, being the only young massive
extragalactic cluster which is not consistent with SSP models when one
includes the effects of rapid residual gas expulsion on the dynamics
of young clusters.

Using high-resolution imaging, MGV05 have reported a change in the
effective radius as a function of wavelength, consistent with the
presence of a large amount of mass segregation.  Using the same data,
Bastian \& Goodwin~(2006) found an excess of light at large radii in
the luminosity profile of the cluster, consistent with the idea that
the cluster is not in virial equilibrium.  However, Goodwin \&
Bastian~(2006) have shown that M82F is extremely sub-virial (i.e. the
velocity dispersion is much too low for the estimated luminous mass
present) if it has a normal IMF, the opposite as would be expected for
residual gas expulsion.

A possible solution to the above discrepancies would be that M82F is younger than presumed in the previous studies, i.e. closer to $\sim10$~Myr than to $60$~Myr.   The age of a cluster,  if assumed to consist of a simple stellar population, can be determined through comparison of the photometry or spectrum with that of models.  Additionally, substructure within a cluster can also be taken as an indicator of the cluster being young  ($<$10~Myr,  e.g. Gutermuth et al. 2005, Elmegreen 2006).  Due to the proximity of M82 ($\sim3.6$~Mpc; Freedman et al.~1994), and the high-resolution capabilities of the Advanced Camera for Surveys (ACS) onboard HST, it should be possible to detect substructure within the cluster (the inner radius of 3~pc is covered by $\sim$144 pixels), hence putting another constraint on the age of the cluster.

Is the IMF in M82F truly different?  In order to answer this question,
strict limits must be placed on all the ingredients that go into the
$L/M_{dyn}$ diagnostic, namely the age, extinction, and size of the
cluster. In the present work we use high-resolution imaging in
multiple bands
to search for substructure within M82F and its surrounding region.  We
complement this dataset with deep Gemini-North GMOS optical
and integral field spectroscopy to accurately determine the age of the cluster and search for a relation
between the cluster and the  H~{\sc ii}
region in which it appears to be enveloped along our line of sight. This paper is organised as follows.  In \S~\ref{data} we
present the imaging and spectroscopic datasets.  In
\S~\ref{sec:results} we analyse the imaging data to search for
substructure in M82F while in \S~\ref{sec:spectra} we use the observed
spectra to age date the cluster.  In \S~\ref{sec:ifu} we use integral field spectroscopy to study the cluster and its surroundings and in \S~\ref{conclusions} we present our conclusions.

\section{Observations}\label{data}
\subsection{HST/HRC Imaging}

The images used in the current study were taken from the {\it HST}
archive fully reduced by the standard automatic pipeline (bias
correction, flat-field, and dark subtracted) and drizzled (using the
MultiDrizzle package - Koekemoer et al.~2002) to correct for geometric
distortions, remove cosmic rays, mask bad pixels and resample the images to a constant pixel scale of 0.025".  The
observations are presented in detail in McCrady et
al.~(2005) and consist of HRC observations in the F250W, F435W, F555W, and F814W (Prop ID: 9473, PI W.D. Vacca).  For the remainder of the current study we will refer to
these bands interchangeably as UV, B, V, and I, although we emphasise that no
transformations have been applied.  Total exposure times for the UV,
B, V and I images were 10160, 1320, 400, \& 120~s respectively.  The
images were aligned using 
point sources throughout the images as reference calibrators, using the
B-band as the reference image, with the {\it GEOMAP} and {\it
  GEOTRANS} routines in {\it IRAF}.  Transformations are accurate
(RMS) to better than $\sim0.2$ pixels in the V and I images, while it is $0.4$
pixels in the UV images due to the lower flux.  Accurate alignments
were necessary to create precise colour maps (e.g.~Seth et al.~2006).

Photometry was carried out on each pixel separately, using the
zero-points given in Sirianni et al.~(2005).   In order to correct for light losses for each pixel, we applied the wavelength dependent aperture corrections of Sirianni et al.~(2005), extrapolated to a single pixel aperture, to the measured magnitudes.  This resulted in colour changes of $-0.06$ mag in $UV-B$ and $0.22$ in $V-I$.

Additionally we carried out
aperture photometry for M82F  by summing all of the flux within a
radius of 15 pixels.  We then used the same background estimate as was
done to make the colour maps (see next section), namely a ring with
inner and outer radius of 50 and 54 pixels respectively.  We note that
the precise location of the background does not significantly affect any of the
results presented here.  We then corrected the summed fluxes for
aperture losses based on the calibrations presented in Sirianni et
al.~(2005), namely 0.16, 0.11, 0.10, and 0.21~mag in the UV, B, V, and I
filters respectively.

To account for the red-leak apparent in {\it UV} filters we corrected
the F250W flux counts by 20\%.  This value was estimated from the
calibrations provided by Sirianni et al.~(2005).  The red-leak was
corrected in the photometry of the models as well.

\begin{figure*}
\begin{center}
\includegraphics[width=8cm]{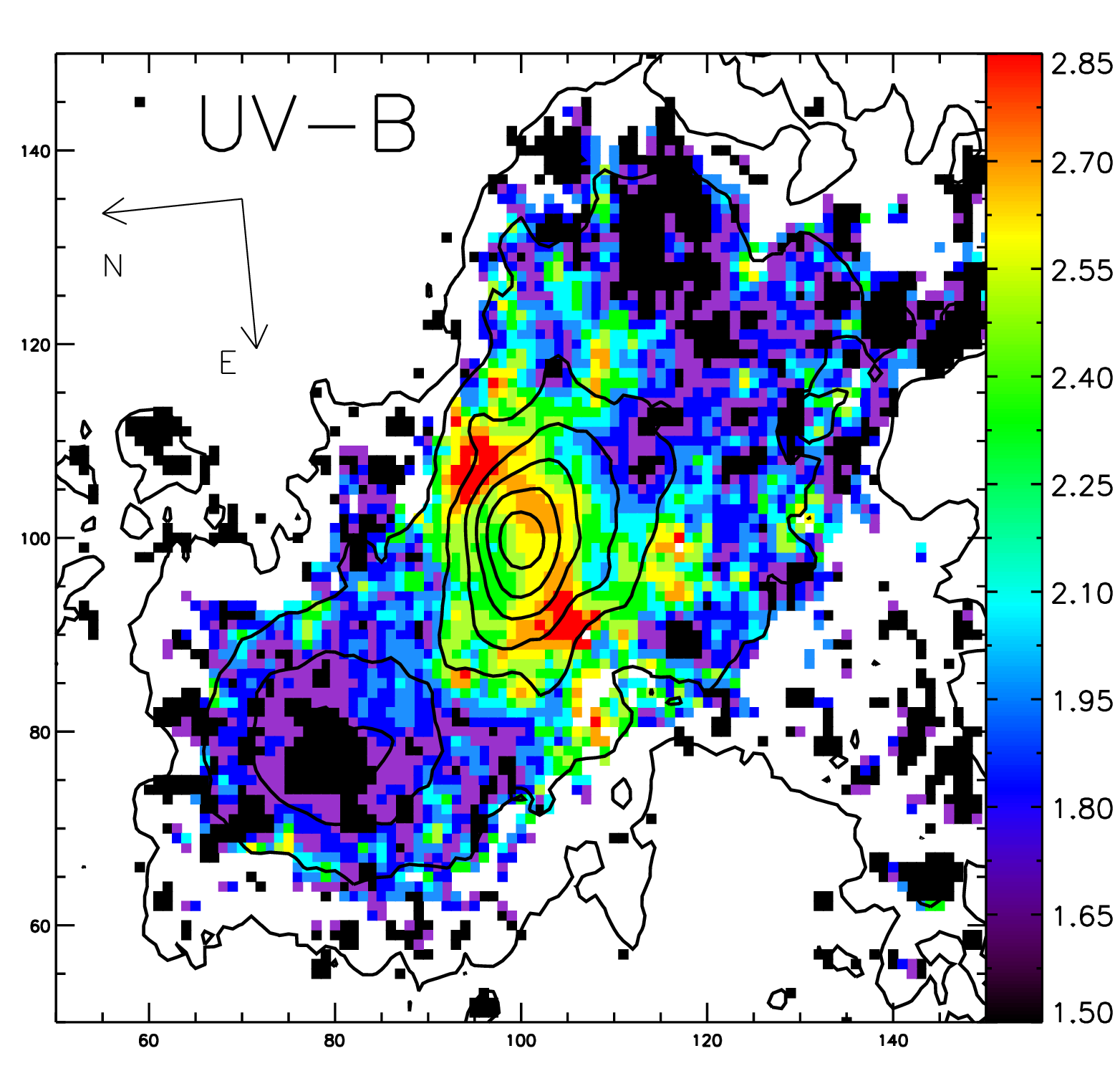}
\includegraphics[width=8cm]{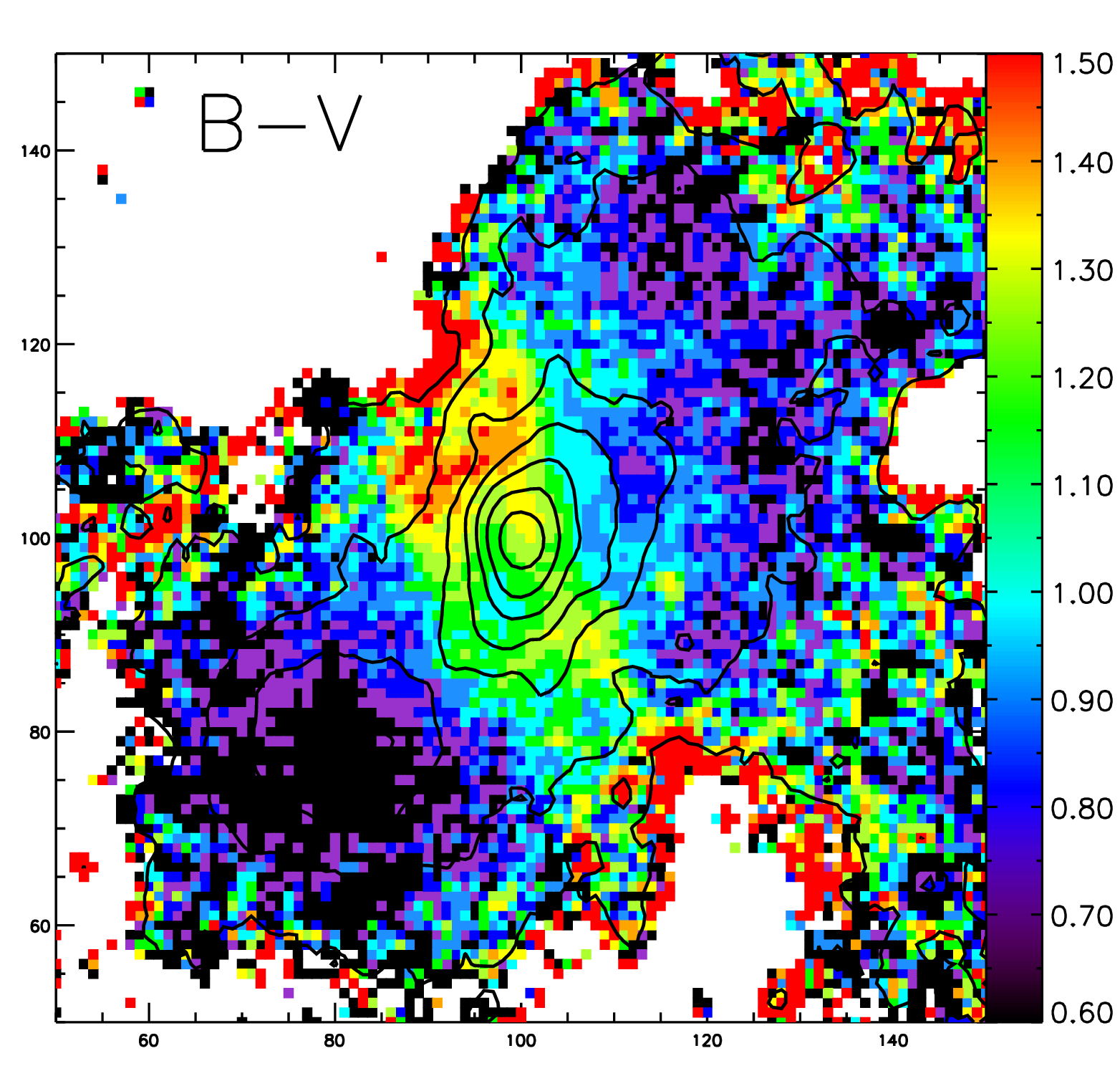}
\includegraphics[width=8cm]{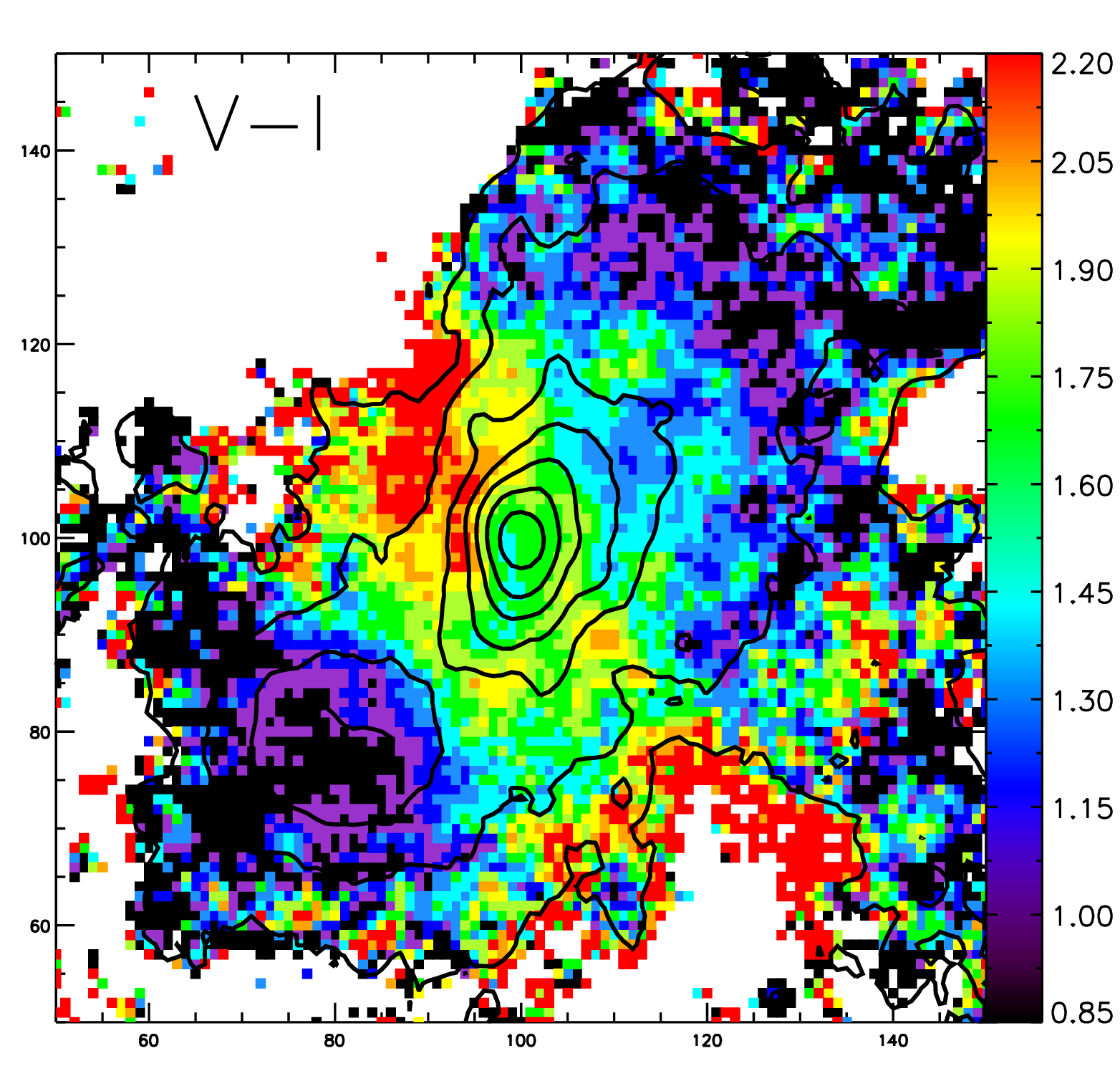}
\includegraphics[width=8cm]{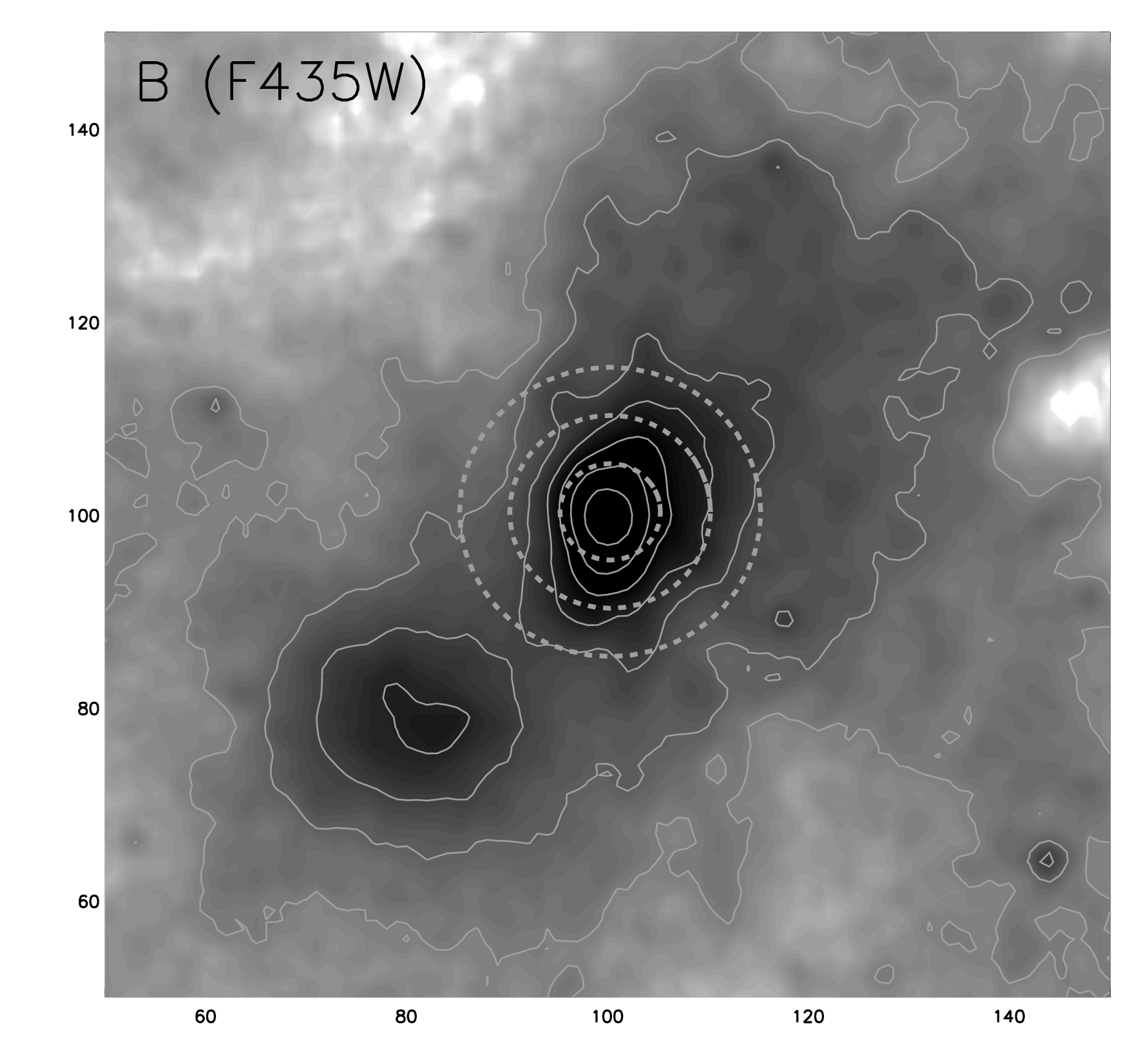}

\caption{{\bf Upper left:} The $UV-B$ colour (F250W--F435W) map of the region
  of M82F.  {\bf Upper right:} The $B-V$ colour
  (F435W--F555W) map of the same region. 
  {\bf Lower left:} The $V-I$ colour (F555W--F814W) map.  {\bf Lower Right:} The logarithmic B-band flux map of the region (no background subtraction has been applied). The three concentric dashed circles show radii of 5, 10 and 15 pixels which are used to colour-code the symbols in Fig~\ref{fig:colour-colour}.  The bars on the right of the three colour maps represent the measured colour in magnitudes.  The contours in
  all panels correspond to the 
  observed B-band flux for 1, 2, 4, 8, 16, 32,  \& 64\% of the peak
  level. All panels have the same orientation and the x and y axes are in pixels (1 pixel = 0.025").  The "companion" cluster is located at (x,y) = (80, 75). }

\label{fig:colour-maps}
\end{center}
\end{figure*}

Once the magnitude of each pixel is known for each band (background
subtracted) we then generate colour maps in order to search for
substructure within the cluster and surrounding region.  These are
shown in Fig.~\ref{fig:colour-maps}.

At the assumed distance to M~82 of 3.6~Mpc, 1 HRC pixel of 0.025" corresponds to 0.44~pc.

\subsection{Spectroscopy}

\subsubsection{Slit spectroscopy}

We supplement the above data set with deep optical spectra taken of
M82F in multi-object spectrograph mode (MOS) using GMOS on
Gemini-North.  The data are part of a larger programme to obtain spectroscopy of the cluster population in M82 (Prog. ID GN-2006A-Q-38; PI L.J. Smith).  The observations were taken on the night of May 4th, 2006 and
consisted of 8 individual exposures of 30 minutes each, for a total
exposure time of 4 hours.  Full details of the reduction 
and data processing will be presented elsewhere.  In brief, we used the B600 grating with a slit width of 0.75" in 2x2 binning mode.   The data were calibrated using a combination of the Gemini IRAF package and custom routines.  The spectra were extracted and wavelength calibrated using arc frames and finally combined to remove cosmic rays.  Flux calibrations were performed using a response function derived from the standard star Wolf 1346.

The resultant spectrum covers
3700\AA--6500\AA~at a resolution of 3.5\AA~at H$\beta$. Additionally,
in Konstantopoulos et al.~(2007, in prep) we will present the full
spectrum of M82F as well as that of 38 other clusters in M82.

\subsubsection{Integral Field Spectroscopy}

In addition to the slit spectroscopy discussed above,                             
we have also obtained Integral Field Unit (IFU)                                   
spectroscopy of M82F and its surroundings using                                   
GMOS on Gemini-North (Prog. ID GN-2007A-Q-                                        
21; PI L.J. Smith). The data were taken on the                                    
night of February 16th, 2007~ in two slit mode. The                                
R831~Êgrating was used to achieve a resolution of 60$\,$km/s and wavelength range of $6150- 6980$\AA.  The field of view is 5 by 7 arcseconds.  Two pointings of                              
900s each were obtained, along with a sky field placed one                        
arcminute away. The orientation of the pointings                                 
was chosen such that the sky field was outside the                               
galactic disk and in a region not contaminated by                                 
the galactic super-wind of M82.Ê     
                                             
The basic reductions of the data were done using a combination of the             
Gemini IRAF package and custom reduction techniques.Ê                             
All science images were bias-subtracted and then                                  
flat-fielded with Gcal (Gemini Calibration Unit) flats and twilight               
flats which were co-added                                                         
and normalized. Spectra were extracted and wavelength calibrated with             
solutions obtained from the arc exposures. The spectra                                      
were cosmic-ray cleaned and the sky was subtracted (using the same techniques as above) and flux                                                    
calibrated using the response function derived from the                           
flux standard star Wolf 1346. As a last step, the two datacubes were                  
resampled to 0.1 arcsec per spatial pixel (spaxel) and combined, which                                       
resulted in a 73x49 spectral cube.

\section{Results}
\label{sec:results}
\subsection{Substructure in M82F}
\label{sec:substructure}

\begin{figure}
\hspace{-0.5cm}
\includegraphics[width=9cm]{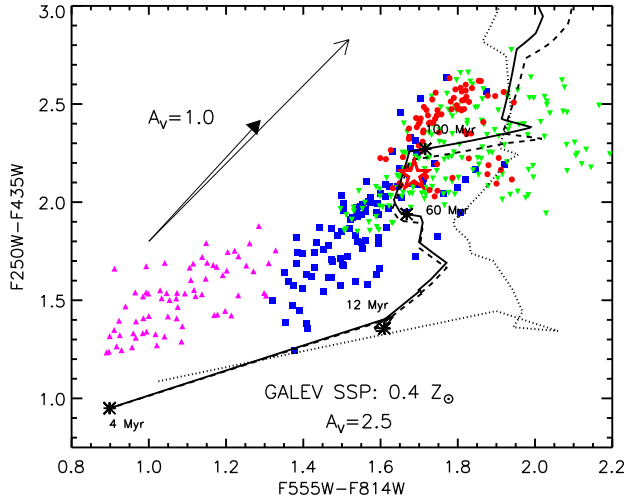}
\caption{The F250W--F435W vs. F555W--F814W ($UV-B$ vs. $V-I$) colour
  plane of the M82F region.  The red circles, green upside-down triangles, and blue squares
  represent pixels located within a 5, 5--10, and 10--15 pixels  radius (2.4,
  4.7, 7.1~pc) of the observed centre of M82F respectively.  The
  magenta triangles represent pixels within 8 pixels of the centre of
  the companion cluster.  The solid line shows the evolution of the
  GALEV SSP models (Salpeter IMF, 0.4~Z$_{\odot}$) reddened by
  A$_{V}$=2.5 magnitudes  (Galactic law). The dashed line shows the same SSP model
  track, except with a lower mass cut-off of 2~$M_{\odot}$ in the
  stellar IMF.   The dotted line is the same as the solid line, except for a solar metallicty SSP.  The asterisks mark ages of 4, 12, 60 and 100 Myr.
  The open (red) star represents the colour of M82F, summed in a 15
  pixel aperture and corrected for aperture losses.  Finally, two extinction vectors are shown as arrows, Galactic extinction law (longer arrow - open head) and the Calzetti extinction law (shorter arrow - filled head).} 
\label{fig:colour-colour}
\end{figure}

In Fig.~\ref{fig:colour-maps} we show the colour maps of $UV-B$,
$B-V$, and $V-I$ for M82F and the surrounding region.  Solid contours
show the observed B-band flux.  Star clusters
are generally assumed to be simple stellar populations therefore we would expect
each pixel to have the same colour\footnote{In practice we would
  expect some scatter as each pixel probably does not fully sample the
mass function of stars.  The scatter perpendicular to the extinction
vector of Fig.~\ref{fig:colour-colour} is presumably due to the
combination of this
effect and measurement errors.}.  However, it is clear that there
is considerable substructure within M82F, which is present in all the colour
maps. Often, this small scale structure is also distinguishable in the observed flux
levels.

In particular we note regions of redder colour to the north-west and south-east of the center of the cluster.  These structures are  clearly visible in the $UV-B$ image (top-left panel of Fig.~\ref{fig:colour-maps}) and also present in all the colour maps.  
The cluster displays bluer colours as one moves away from the cluster center to the south-west and north-east.  We note that due to the lower flux level in the UV image, the $UV-B$ colour map does not extend as far to the north-west as the $B-V$ and $V-I$ maps, hence the `red' region to the north-west of the later images does not appear in the $UV-B$ map.  We find substructure throughout the projected face of the cluster in all of the colour maps shown in Fig.~\ref{fig:colour-maps}, along with good agreement in the position, shape and relative intensity of the substructures in each of the maps.

In order to search for the origin of this substructure we show in
Fig~\ref{fig:colour-colour} the 
($UV-B$)~vs.~($V-I$) colour distribution for each pixel in M82F and the
surrounding area.  We also show two
extinction vectors for $A_V=1$~mag, where the longer arrow (open head) is for a standard Galactic extinction law of Savage \& Mathis~(1979) and the shorter arrow (filled head) represents the Calzetti~(1997) starburst extinction law. Also plotted as a solid line we show the {\it GALEV} (Anders \& Fritze v. Alvensleben 2003) SSP  model tracks
of Salpeter IMF, $0.4~Z_{\odot}$ metallicity, which use the Padova stellar isochrones. These tracks have been shifted
by $A_V=2.5$~mag assuming the Galactic extinction law. The
dashed line shows the same SSP models (also shifted by $A_V=2.5$~mag),
but now for a lower mass 
cut-off in their stellar IMF at $2~M_{\odot}$ (similar to the best
fitting IMF found by Smith \& Gallagher~2001).  

The integrated colour of M82F is shown as a open (red) star, which has been corrected for aperture losses.  The colours of M82F are
consistent with an age of 60~Myr and an extinction value of
$A_V=2.8$~mag, as found by Smith \& Gallagher~(2001).  We note that the solar metallicty SSP models (dotted line in Fig.~\ref{fig:colour-colour} do not reproduce the colours of M82F as well as the $0.4~Z_{\odot}$ models.  However, some care must be taken not to over-interpret this result,  as the aperture corrections applied to the photometry were derived from point sources whereas the cluster is clearly resolved.  The true cluster integrated colours may be somewhat redder (especially in $V-I$) than shown here.  This being the case, we cannot distinguish between the $0.4~Z_{\odot}$  or 1~$Z_{\odot}$ model tracks, although ISM studies of M82 suggest values closer to the solar value (McLeod et al.~1993), similar to that found from near-IR spectra of red super-giants (Origlia et al.~2004).  However, since the SSP model tracks are parallel at these ages the choice of the model does not affect our results.  Throughout the remainder of the paper we will discuss the interpretation for both metallicities.

The fact that the pixels in M82F mainly scatter parallel to the extinction vector lends strong support
to the notion that the substructure seen in the colour maps is due to differential extinction across the cluster.  While some of the substructure seen in the colour maps may be due to differential ages (i.e. the extinction vector and SSP models as a function of age are somewhat parallel in the observed regime), the age derived by spectroscopy argue for an older age, namely $\sim60$~Myr  (e.g. Gallagher \& Smith~(1999), and \S~\ref{sec:spectra}).  Age differentials in such a small region as a cluster are expected to be much smaller, on the order of a few Myr (e.g. Elmegreen 2006), thus ruling out differential ages as a major contributor to the scatter seen in Fig.~\ref{fig:colour-colour}.

\subsection{Extinction across M82F}
\label{sec:extinction}

Assuming that the substructure presented in
\S~\ref{sec:substructure} is indeed due to differential extinction, then it
is possible to construct an extinction map for the region.  To do
this, we assume that each pixel has the intrinsic colour of an SSP
model with an age of 60~Myr (see \S~\ref{sec:spectra}).  We then shift each observed pixel value 
along the extinction vector in $B-V$ vs $V-I$ space (the UV band was
not used in this analysis as the flux per pixel was too low in
the outer regions) until the value closest to the intrinsic colour is
reached.  The resulting extinction map is shown in the top panel of
Fig.~\ref{fig:ext-map}.  Again, the solid contours represent the
observed B-band flux.  We only show the region within 30 pixels from the
observed centre of M82F, as outside this radius it is unclear what the
intrinsic colour of each pixel should be.  For this we used a background
region identical to that used to make the colour maps.   We note that
choosing a different age (e.g. 10~Myr) for the intrinsic colour does not
change the content of Fig.~\ref{fig:ext-map} or the subsequent
analysis, as every pixel is then corrected by the same additional factor
in extinction.

A few features stand out clearly in the extinction map (i.e. the top
panel of Fig.~\ref{fig:ext-map}).  The first is a high extinction
region to the north-west of the centre of the cluster.  This feature
is presumably responsible for the tightening of the observed flux
contours with respect to the opposite side where the distribution is
much smoother.  Also, on the opposite side (to the south-east) there
seems to be a small patch of high extinction which is also reflected
in the skewness of the observed flux contours.

Additionally, there exists a lane of high extinction which connects
these two regions, passing through or near to the observed cluster
centre. There also appears to be a ``hole" in the extinction map to the north-east of the cluster.  

We note that adopting a different interstellar extinction law does not substantially affect our results.  For example, if we used the extinction law of Calzetti~(1997) all of the structures seen in Fig.~\ref{fig:ext-map} are still present,  with their relative strengths preserved.  However, the total amount of extinction for  every pixel would increase by approximately $\Delta$A$_{V}$=0.9~mag.

While the south-west side of the cluster also shows substantial
differential extinction, the total amount of extinction is much less.
On the opposite side,  it is not clear if the low extinction values
represent intrinsic values as the colour of the companion cluster is
much bluer (shown as magenta triangles in Fig.~\ref{fig:colour-colour}), which may simply represent a younger age.   The much lower extinction estimated for the 'companion' cluster suggests that this cluster is not physically associated with M82F, but instead resides closer to the observer.  This cluster, seen in the images and colour maps at (x,y) = (80,75),  does show some evidence for differential extinction across its face, however it also seems offset from the colour of M82F.  This suggests that the cluster may be younger and perhaps associated with the H~{\sc ii} region.  Unfortunately, no information exists on the velocity of this cluster (see \S~\ref{sec:velocity} \& \S~\ref{sec:ifu} for a discussion of the surrounding H~{\sc ii} region).

Upon close inspection, one difference between Fig.~\ref{fig:colour-colour}  and Fig.~\ref{fig:ext-map}
can be seen, namely that the highest extinction region to the north-west of cluster centre is not present in the colour-colour plot.  The reason for this is the inclusion of the UV in Fig.~\ref{fig:colour-colour}, which was excluded (due to the lower flux levels) in the derivation of the extinction.  This has been checked with $B-V$ vs $V-I$ plots, which themselves do not enable us to distinguish between age and extinction, but which can be used to derive the extinction if the age is already known.

In Table~\ref{table:info} we list the derived extinction and absolute magnitude of M82F.  The fact that the dust lanes seen on and around M82F do not display any connection to the cluster's structure, argues that the dust causing the extinction is in a 'screen', i.e. not intermixed with the stars of the clusters.

\subsubsection{The effect of a wavelength dependent PSF}

Since {\it HST} produces near diffraction limited images, the PSF of the HRC is wavelength dependent (e.g. Sirianni et al.~2005).  To test whether this biases our results we have performed the following tests.  We have smoothed each image with a gaussian kernel with $\sigma=$1, 2, and 3 pixels.   Tests using the PSF images showed that smoothing with the largest value (i.e.~$\sigma=3$~pixels) effectively removed any structures in colour maps of the PSF.  We then generated the extinction maps using these smoothed images and searched for differences between these maps and the unsmoothed map.  Even with the lower resolution, all of the major structures discussed above are still present, namely the high extinction regions to the north-west and the south-east of the centre of the cluster.

Additionally, we can use the colour images in Fig.~\ref{fig:colour-maps} to search for PSF effects.  The PSF for the F435W and the F555W are very similar to each other, compared with the F814W PSF (Sirianni et al.~2005).  Furthermore, the F250W and F435W have very similar PSFs.  Therefore we expect to see the same sub-structure in the UV--B and B--V images as in the derived extinction map.  In fact, this is what we find.  Again the two high extinction regions (to the north-west and south-east) of the center, along with a general filament across the cluster can be seen as redder colours in both of these colour maps.  Additionally, we see the small 'hole' in the extinction to the north-east of the cluster centre.

Thus we conclude that the wavelength dependence of the PSF is not significantly affecting the results presented here.

\subsection{The intrinsic shape of M82F}

\begin{figure}
\begin{center}
\includegraphics[width=8.5cm]{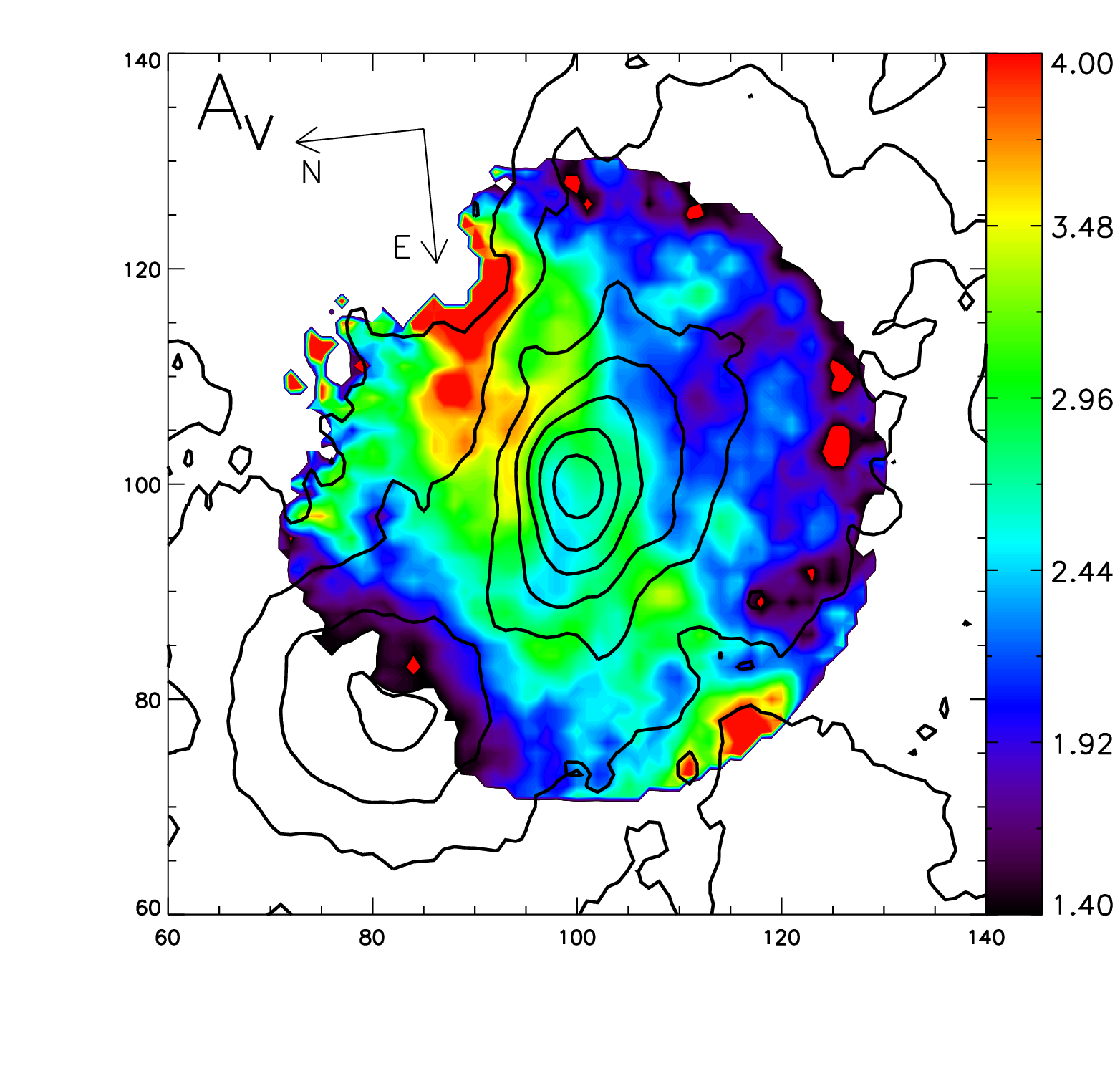}
\includegraphics[width=7.5cm]{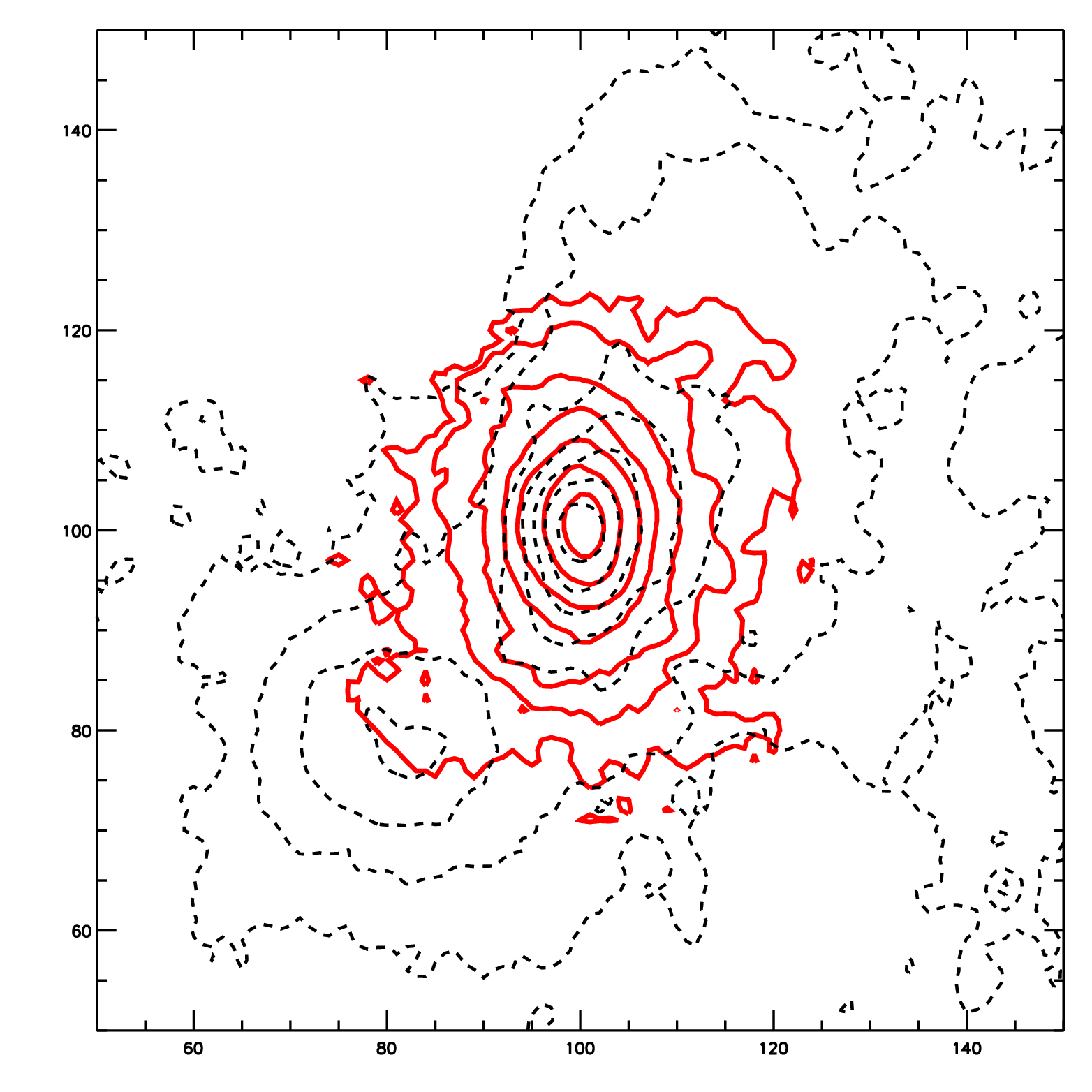}
\caption{{\bf Top:} The colours represent the extinction map of the
  region around M82F.  Note the strong spatial
  dependence of the extinction. The solid contours represent the
  observed B-band flux.  The bar on the right of the panel shows measured $A_V$ in magnitudes. {\bf Bottom:}  The solid (red) contours show
  the B-band flux after correction for the derived extinction (from
  upper panel).  The dashed contours represent the observed B-band
  flux.  The contours represent the $1, 2, 4, 8, 16, 32,$ \& $64\%$ flux levels of the maximum flux in the intrinsic and observed images respectively.
  The orientation is the same as in the top panel and in Fig.~\ref{fig:colour-maps}  Both panels are given in pixels, where one pixel corresponds to 0.025".}
\label{fig:ext-map}
\end{center}
\end{figure}

Using the derived extinction map shown in the top panel of
Fig.~\ref{fig:ext-map}, we can correct the observed flux per pixel for
reddening, and derive the {\it intrinsic} flux distribution
(referred to as the intrinsic image) of M82F.  This is shown in the bottom panel of
Fig.~\ref{fig:ext-map}, where the intrinsic B-band flux contours are shown as
solid (red) lines and the observed flux contours are shown as dashed
(black) lines.  Both sets of contour levels are shown as $1, 2, 4, 8,
16, 32,$ and $64\%$ of their maximum values.  The companion cluster, at (x,y) = (80, 75), still apears in the intrinsic image (as an extended lobe to the north-east of the centre of M82F), although its presence is somewhat weaker than in the original images.  This is due to its bluer colour, which when Fig.~\ref{fig:ext-map} is produced, is assumed to reflect a lower extinction.  However, as can be seen in Fig.~\ref{fig:colour-colour} the companion does seem shifted in colour space, suggesting a younger age.  Thus the corrected image in this region is more uncertain and will not be used in the following analysis.

As a test of the significance of the derived extinction and contour maps we have added Poissonian noise to the observed images and carried out the same analysis as given above.  We did not find any noticeable differences in the derived maps, thus we conclude that the main features seen in the extinction map and the subsequent extinction-corrected image (contour map) are real.

In order to search for structural differences between the observed and
intrinsic images, we used {\it ISHAPE} (Larsen~1999) to fit King
(King~1962) and
Elson, Fall, \& Freeman~(EFF, 1987) profiles to the observed flux distributions.  The King and EFF profiles feature similar central cores, however they differ in the outer regions, with the King profiles having a distinct truncation at large radii while the EFF profiles follow an asymptotic power-law fall-off without any such truncation. 

For all experiments we use a fitting radius of 15~pixels.  We have generated the PSF for each filter in two different ways.  First, we have used the programme {\it TinyTim} to generate a PSF at the position of M82F on the chip.  Secondly, we have generated an empirically derived PSF based on observations of resolved star clusters in the galaxy and LMC.  We do not find any systematic differences between the results of the two PSFs.

For all experiments we use a fitting radius of 15~pixels.
Fitting on concentration parameters for King profiles leads to values
of $8.5-11.5$ for both the intrinsic and observed profiles. We found
$\chi^2$ values of the fits to be 4 times lower for the intrinsic
images for the B and V bands, while they were slightly higher for the I
band,  compared to the observed images.  The best fitting effective radius (corrected for the
ellipticity) is R$_{\rm eff}$=0.12, 0.12 and 0.11 arcseconds for the
observed profile of the B, V, and I bands, in good agreement with that
found by MGV05, namely $0.120\pm0.002$, $0.123\pm0.002$, \& $0.113\pm0.002$ (from their Fig.~4).  However for the intrinsic profile we
find R$_{\rm eff}$=0.13, 0.12 and 0.11 for the B, V, and I bands
respectively.  

For young clusters King profiles may not be the best representative function.  Young clusters tend not to have a tidal truncation, but instead are best fit by an EFF profile. Therefore we have also fit EFF profiles to the images and find that they give equally good fits.  The best fitting index of the EFF profiles is 1.46 or $\gamma = 2.92$.  The effective radius of these fits is $\sim13-20\%$ larger  than that determined using a King profile in both the observed and intrinsic images.  Using an EFF profile on either the intrinsic or observed images appears to eliminate the wavelength dependence of the $R_{eff}$, therefore calling into question previous reports of mass-segregation in M82F (MGV05).  From the EFF profile fits carried out on the intrinsic images we find an average, ellipticity-corrected, effective radius of 0.15 arcseconds, or 2.63~pc ($\pm 0.32$) at the assumed distance of M82F (where the error is the standard deviation between the three filters).   The derived parameters for all filters in both the intrinsic and observed images are given in Table~\ref{table:size}.

In order to test for degeneracies in the fit parameters, we used version 0.93.9 of {\it ISHAPE} which takes into account correlations between the parameters.  As an example (in fact the worst case), we give the best fit parameters along with the 1$\sigma$ errors for the intrinsic F435W image in Table~\ref{table:size}.  While the FWHM, major/minor axis, and index can shift by 7, 9, \& 15~\% respectively, relative to the best fit value, we find that the combination of these in order to derive the effective radius varies by only $\sim10$\%. 

Since there is no evidence for a tidal truncation in M82F (Bastian \& Goodwin 2006), we adopt the effective radius from the  EFF profile rather than the King profile, leading to an adopted size (effective radius) of 2.63~pc for M82F. 

In Table~\ref{table:info} we give the measured size, the velocity dispersion from the literature, and the corresponding light to mass ratio.

The profile of M82F is extremely steep.  Only $\sim17\%$ of known globular clusters in the Galaxy have King profiles with indices ($\log c$) less than 1 (Harris~1996).   Comparing the EFF profile fits of M82F to those of LMC clusters shows that only $\sim22$\% of the clusters there have profiles with $\gamma$ greater than 3 (Mackey \& Gilmore~2003).

\begin{figure}
\begin{center}
\includegraphics[width=8cm]{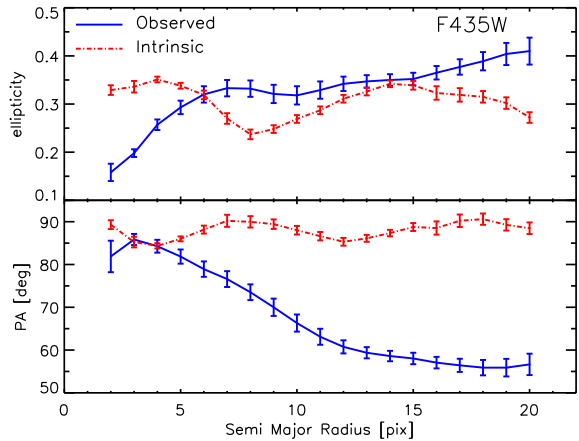}
\caption{{\bf Top:}  The observed (solid blue line) and intrinsic
  (dash-dotted red line) ellipticity of M82F as a function of distance from the
  centre of the cluster. {\bf Bottom:}  The relative position angle
  (same labelling as the top panel) as a function of distance from the centre.}
\label{fig:pa}
\end{center}
\end{figure}

Following on these results we also used the {\it IRAF} routine {\it
  ELLIPSE} to measure the ellipticity and position angle of the
  cluster as a function of distance from its centre.  The results for
  the B-band are shown in Fig.~\ref{fig:pa}.  In the top panel we show
  the ellipticity of the cluster for both the observed (blue solid)
  and intrinsic (red dash-dotted) images.  Note that the ellipticity
  is highly dependent on radius for the observed cluster profile (varying from
  0.15 to 0.4 in the inner 20~pixels), while the intrinsic profile
  varies much less ($0.3\pm0.05$) in the same interval.  In the
  bottom panel of Fig.~\ref{fig:pa} we show the position angle for the two cluster
  profiles as a function of radial distance.  There is a strong radial trend ($\sim25$~degrees) in the
  observed profile while the intrinsic profile does not show any
  systematic trend with radius and the amplitude of variation is much
  lower ($\sim7$~degrees).  This figure agrees well with the results
  from {\it ISHAPE}.

\subsection{The high L/M$_{\rm dyn}$ of M82F}

As shown above the variable extinction across the face of M82F complicates in the determination of the clusters intrinsic luminosity and size.  Both of these quantities are essential in determining the \lm (ratio of the intrinsic luminosity to its dynamical mass).  By comparing this ratio to that of SSP models one can place constraints on the underlying mass function of the cluster (e.g. SG01, MGV05, Bastian et al.~2006) or on its dynamical state (e.g. Goodwin \& Bastian~2006).

In Table~\ref{table:info} we show the measured intrinsic luminosity and effective radius and the derived \lm~ ratio using the velocity dispersion ($\sigma$) measured by SG01 and the virial theorem (for details see SG01 and MGV05). 
For a 60~Myr simple stellar population, the expected \lm~lies in the range of 
$\sim7 - \sim12~L_{\odot}/M_{\odot}$ for solar metallicity and a Salpeter (1955) or Kroupa (2002) initial stellar mass function (Maraston 2005).  The derived value for M82F is 42 ($\pm 10$) ~$L_{\odot}/M_{\odot}$, hence it is well outside the range expected for a standard IMF, as has been previously noted by SG01 and MVG05.  If the cluster is out of dynamical equilibrium due to either the expulsion of natal gas (Goodwin \& Bastian 2006) or an interaction with a giant molecular cloud (Gieles et al. 2006) we would expect the \lm~ to be {\it lower} than that of the SSP model for the correct age.

Thus, even having taken into account differential extinction across M82F, the large value of \lm~ is still not accounted for.  Possible alternatives are a truly top-heavy stellar initial mass function, a highly mass-segregated cluster (MGV05), or unforeseen complications in the velocity dispersion measurements due to the large differential extinction.

\section{Results from Deep Optical Spectroscopy}
\label{sec:spectra}

\subsection{Age}

Due to the complexity of M82F there has been some debate in the
literature as to the precise age of the cluster.  SG01 have assigned
an age of 60~Myr based on the width (FWHM) of the optical Balmer lines and
the depth of the He~I lines while MGV05 have derived similar age, namely between $40-60$~Myr, based
on their H-band spectra. However, SG01, Bastian et al.~(2006b) and Goodwin
\& Bastian~(2006) have noted that the reported discrepancy between the L/M$_{\rm dyn}$ ratio of the
cluster and SSP models would be resolved if the cluster was less than
$\sim15$~Myr old.  Bastian \& Goodwin~(2006) have provided some
evidence for this scenario by showing that the cluster profile is in a
non-equilibrium state, suggestive of a young age.  Additionally the
photometry (i.e.~Fig.~\ref{fig:colour-colour}) allows for multiple
solutions to the age of M82F (one at $\sim60$~Myr and the other around
10~Myr).  Therefore we have
obtained high signal-to-noise (S/N $>~200$ per spectral pixel) optical spectra of this cluster in order
to help resolve the debate.

In order to find the age of M82F, we have compared the observed
spectra to SSP models, taken from Gonz\'alez Delgado et
al.~(2005, hereafter GD05),  where we have chosen to use the Padova
isochrones in order to provide the best comparison with the GALEV models used in the photometric analysis.  Additionally, we have also chosen the models with a Salpeter~(1955) stellar IMF (in the range of 0.1 to 120~$M_{\odot}$) which was also used in the adopted GALEV models.  Finally, all results will also be discussed in terms of solar and 0.4~solar metallicities.

\begin{figure}
\begin{center}
\includegraphics[width=8cm]{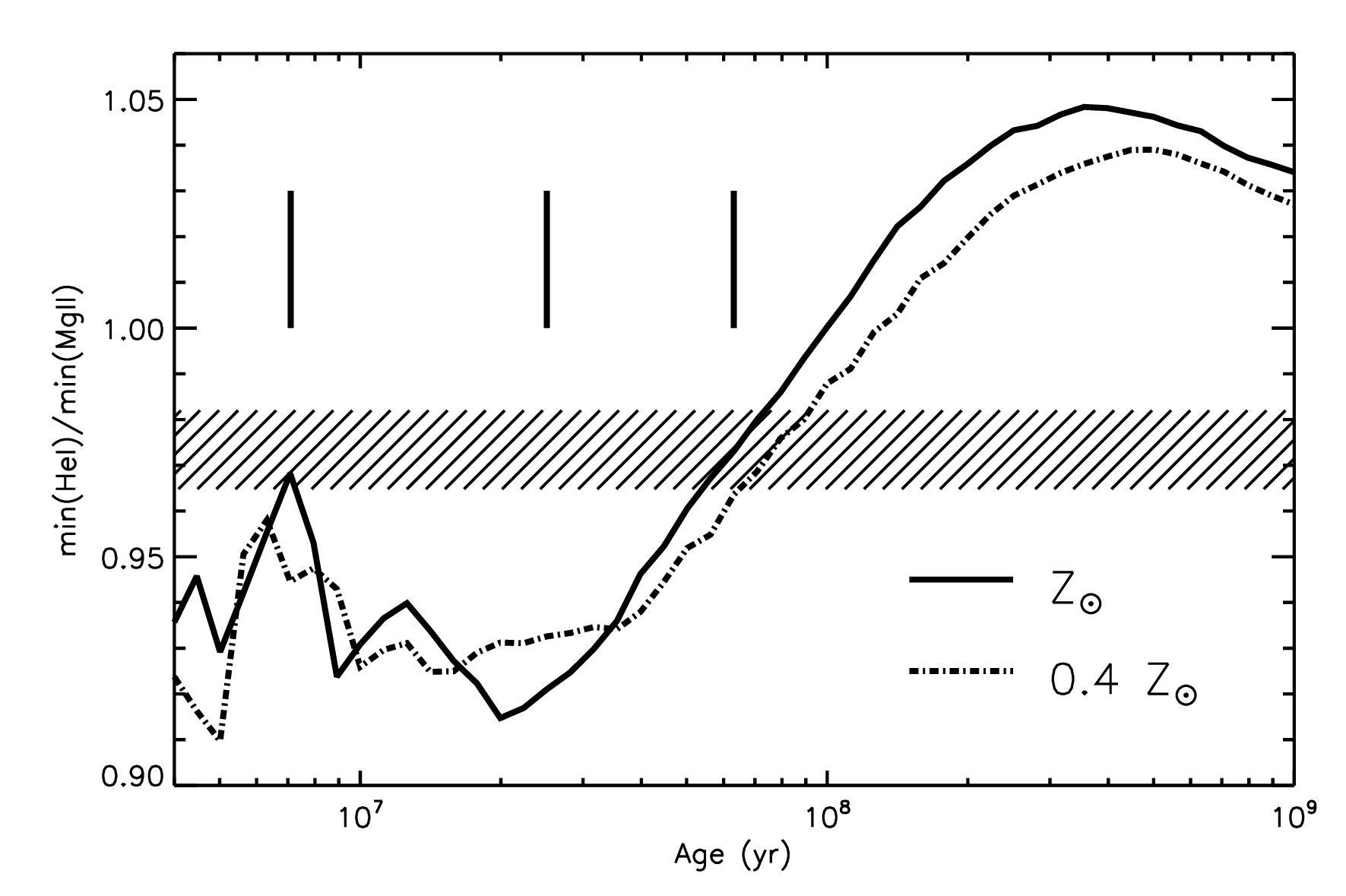}
\includegraphics[width=8cm]{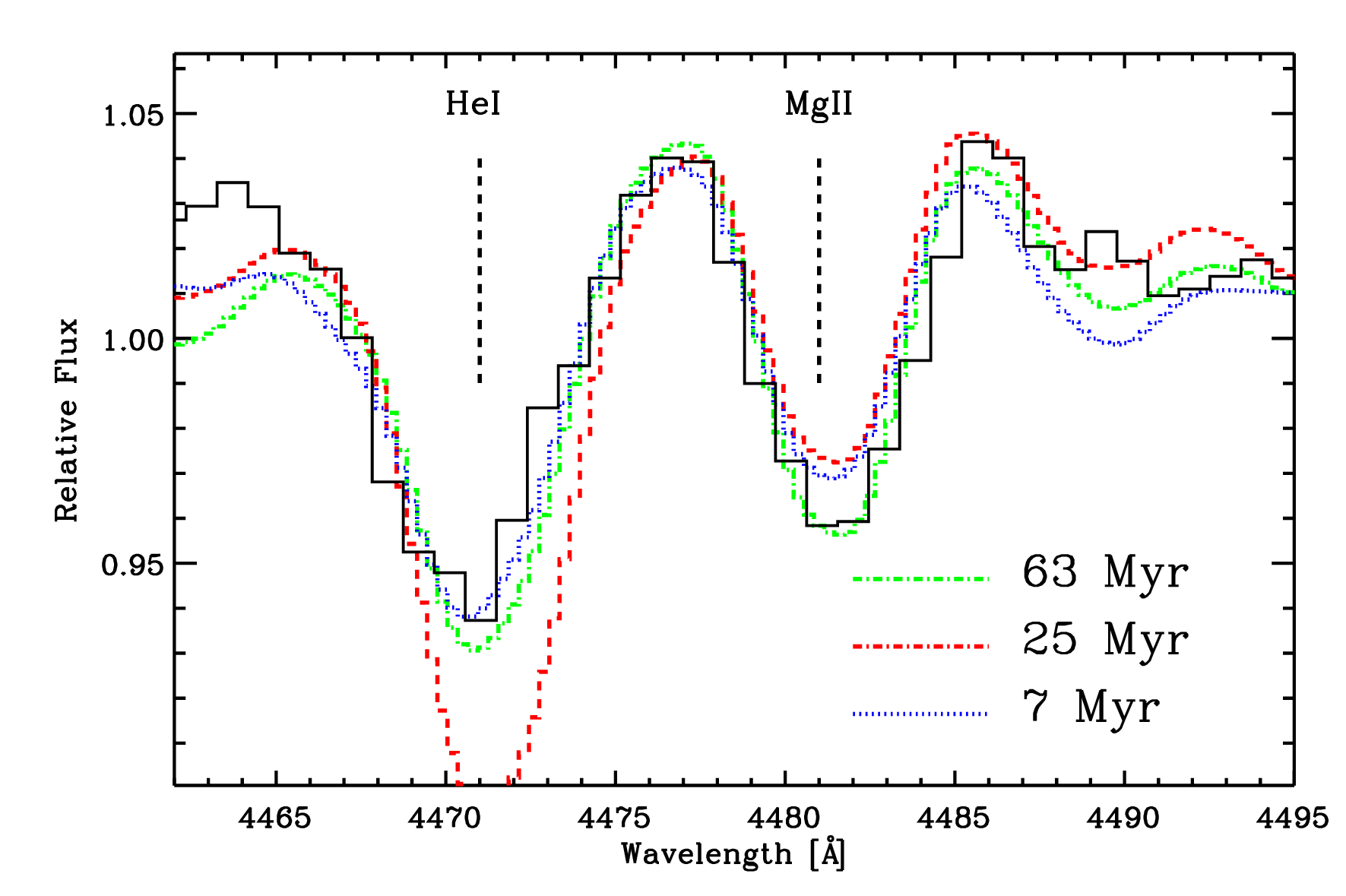}
\includegraphics[width=8cm]{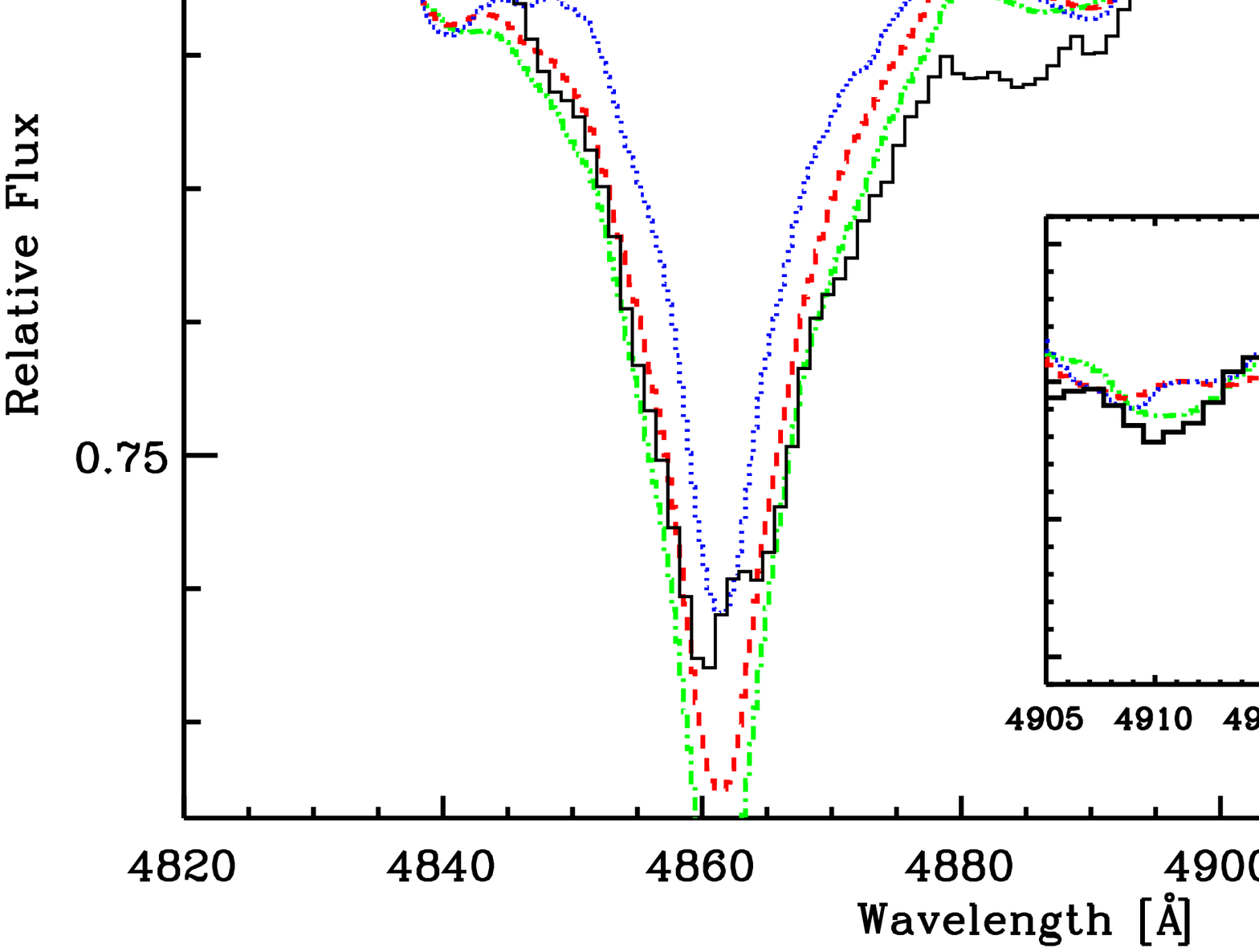}
\caption{{\bf  Top panel:}  Prediction from SSP models as to how the
  ratio of the depth of He~{\sc i}\,$\lambda4471$ and Mg~{\sc ii}\,$\lambda4481$ changes
  with age.  Two metallicities are shown (see text for details of the
  models).  The hashed region is the measurement for M82F including the
  effects of extinction and noise in the data.  The vertical lines
  mark the age of the models shown in the middle and bottom panels. {\bf Middle
  panel:}  The region of the spectrum around the He~{\sc i} and Mg~{\sc ii} lines.
  The black line represents the observations while the other lines
  indicate model spectra for three different ages labelled on the
  figure. {\bf Bottom panel:}  Region of the spectrum around H$\beta$
  and He~{\sc i}\,$\lambda4921$. Note that the younger models do not fit the
  width of the Balmer lines, nor the depth of the He~{\sc i} line.}
\label{fig:spectra}
\end{center}
\end{figure}

 All SSP models used in this work have been smoothed  to have
 approximately the same resolution as the data.   Due to the high
 extinction of M82F (A$_{\rm V}=2-4$) the best age determining
 mechanism in the optical would be the ratio of two close strong lines
 (in order to avoid problems with extinction) which are highly
 sensitive to the temperature of the stars which produce them.  Such a
 ratio is provided by the He~{\sc i}\,$\lambda4471$ and Mg~{\sc ii}\,$\lambda4481$
 lines.  The ratio of the depths of these two lines as a function of
 age of the SSP model (for solar and 0.4 solar metallicity) is shown
 in the top panel of Fig.~\ref{fig:spectra}. The depth of each line, i.e. the minimum, is found in two ways.  The first is simply measuring the flux value at the minimum of each line, which, however, may be subject to Poissonian noise.  The second method is to fit each line with a Gaussian function and take  the minima.  For the observed spectrum, both methods were employed on the original as well as the extinction corrected spectra (corrected values between $A_{V}=0-4$~mag).  The hashed area in the top panel represents the observed ratio of M82F for both methods and all extinctions tested.
 
 As {demonstrated by the top panel of Fig. \ref{fig:spectra}}, the relative
 depth of these two lines is a good age indicator for ages between  
$\sim40-400$~Myr.  The middle panel
of this figure shows the region of the spectra around the He~{\sc i} and Mg~{\sc ii} lines between
$\lambda\lambda$4465--4490\AA~ with three SSP models of different ages superposed. In order to remove the effects of extinction and paralactic angle on the observed spectra, we attempted to rectify the spectra shown in the middle and bottom panels of Fig.~\ref{fig:spectra}.  To do this we first divided each spectrum by the mean flux in two adjacent continuum bands and then fit a second order polynomial to the continuum bands on each side of the line of interest (this was done independently for the He~{\sc i}/Mg~{\sc ii} lines and the hydrogen Balmer series).  The fit was then subtracted from the observed spectra.  
 
Some care must be taken when using the He~{\sc i} to Mg~{\sc ii} ratio as emission could be present in the He~{\sc i} line, due to the presence of ionised gas (SG01).  If there is a significant amount of emission within the He~{\sc i}\,$\lambda4471$ line, the He~{\sc i}/Mg~{\sc ii} ratio would be pushed to lower values, biasing the age determination to lower values.  This does not seem to be a problem in the present case, as no emission is found for He~{\sc i}$\,\lambda5876$ which would show the largest nebular contribution.

From this and assuming solar metallicity we can conclude that the age of M82F is between 50 and 70~Myr, in good agreement with
the values found by GS99, SG01, MGV05. If the metallicity of M82F is similar to that of
the LMC then the derived age of M82F would be older by
$\sim10-15$~Myr.  However, as seen in the top panel of
Fig.~\ref{fig:spectra} a younger age of 7~Myr is also consistent with
the data.  This age can be ruled out by the width of the Balmer lines
and depth of the Helium lines which are predicted to be much narrower
and deeper respectively for young stellar populations (bottom panel of
Fig.~\ref{fig:spectra}). The fact that
the SSP models over-predict the depth of H$\beta$ is indicative of the
presence of emission within the line.

\begin{table}
{\scriptsize
\parbox[b]{8cm}{
\centering
\caption[]{The properties of M82F.}
\begin{tabular}{c c c c}
\hline
\noalign{\smallskip}
Parameter  & Value \\
\hline
Age & 50--70~Myr \\
R$_{\rm eff}$ & 2.63 $\pm~0.32$ pc \\
A$_{V}$ & 2--4 mag \\
M$_{\rm V}$ & --14.3 $\pm~0.2$ mag$^a$\\
$\sigma$ & 13.4 $\pm$~0.7 km/s $^b$\\
L/M$_{\rm dyn}$ & 42 $\pm~10$ L$_{\odot}$/M$_{\odot}$\\

\noalign{\smallskip}
\noalign{\smallskip}
\hline
\end{tabular}
\begin{list}{}{}
\item[$^{\mathrm{a}}$] Using an aperture of 20~pixels and the background region defined in the text.  
Measured from the intrinsic V-band image (i.e. extinction corrected image).
 \item[$^{\mathrm{b}}$] Taken from SG01.
\end{list}
\label{table:info}
}
}
\end{table}

\begin{table}
{\scriptsize
\parbox[b]{8cm}{
\centering
\caption[]{EFF profile fits to M82F.}
\begin{tabular}{c c c c c c}
\hline
\noalign{\smallskip}
Filter  & FWHM$^{a}$ (pix) & e$^{b}$ & Index$^{c}$  & R$_{\rm eff}^{d}$ (pc)\\
\hline
{\bf Observed}  \\
F435W & $6.75$ & $0.58$ & $1.65$ &$2.22$ & \\
F555W & $5.11$ & $0.56$ & $1.35$ &$2.65$ &\\
F814W & $5.70$ & $0.53$ & $1.59$  & $1.93$\\
Average$^{e}$ & & & &  $2.25\pm0.33$\\
\hline
{\bf Intrinsic} \\
F435W &  $6.99^{+0.48}_{-0.19}$ & $0.63^{+0.06}_{-0.05}$ & $1.58^{+0.28}_{-0.06}$ & $2.54^{+0.04}_{-0.26}$ \\
F555W &  $5.56$ & $0.56$ &  $1.34$ & $2.99$\\
F814W &  $5.71$  & $0.55$ & $1.45$&  $2.36$\\
Average$^{e}$ & & & & $2.63\pm0.32$\\
\noalign{\smallskip}
\noalign{\smallskip}
\hline
\end{tabular}
\begin{list}{}{}
\item[$^{\mathrm{a}}$]   FWHM of the major axis.
\item[$^{\mathrm{b}}$]   The FWHM of the minor axis divided by the FWHM of the major axis.
\item[$^{\mathrm{c}}$]   The index of the Moffat profile, equivalent to $\gamma/2$ of an EFF profile.
\item[$^{\mathrm{d}}$]   The effective radius of the best fitting profile, correcting for ellipticity and using the conversion between the FWHM and R$_{\rm eff}$ for the best fit value of the index. 
\item[$^{\mathrm{e}}$]  The average and standard deviation of the best fit results for the three different filters. 
 \end{list}
\label{table:size}
}
}
\end{table}

\subsection{Velocity}
\label{sec:velocity}

In order to find the velocity of M82F we have cross-correlated the
observed spectra (extinction corrected by A$_{\rm V}$=3~mag) with the GD05 SSP templates.  For the cross-correlation we have used two methods, the {\em IRAF} task {\em FXCOR} and the Penalized
Pixel Fitting (PPxF) routine of Cappellari \& Emsellem~(2004).  In
both techniques we have masked out emission lines and instrument
defects such as chip gaps in the observed spectra.  Both methods
gave consistent results with the heliocentric velocity of the
absorption component (presumably the cluster) of 15.9
($\pm$~15)~km/s.  This result is in good agreement with SG01 who found a
velocity of 24.3 ($\pm$~1.7)~km/s using high resolution spectra.

The PPxF method allows one to subtract the best fitting template in
order to extract the information on the emission lines present.  In
the present case, emission from H$\beta$ and [OIII]$\lambda$5007\AA~is
clearly detected.  The heliocentric velocity of this component is
40.8 ($\pm$~10.3)~km/s .  This is also consistent with the first
emission component found by SG01 at 50~km/s but we do not see strong
evidence of the components at 120 and 180~km/s.  

The velocity difference between the emission and absorption
components, along with the derived age, strongly suggest that M82F is
not physically associated with the H~{\sc ii} region which surrounds it in
projected images.  This will be discussed further in the next section.

\section{IFU studies of M82F and its surroundings}
\label{sec:ifu}

 Integral field spectroscopy of young stellar clusters offers an ideal tool to study their effect on the surrounding ISM and other nearby clusters (e.g. Bastian et al.~2006, Westmoquette et al.~2007, in prep).  As discussed above, M82F is located in a complex region of the galaxy and appears projected near a massive star forming site, whose emission lines appear superposed on the broad absorption lines of the cluster.  In order to further investigate where precisely this emission is coming from and how it relates to M82F we have obtained an IFU pointing, centred in wavelength space on H$\alpha$ and the [N~{\sc ii}]\,$\lambda6583$~\AA~ (hereafter [N~{\sc ii}])~ lines, although the [S~{\sc ii}]\,$\lambda\lambda6717, 6731$ lines are also covered.  Figure~\ref{fig:ifu-pointing} shows the position of the IFU field of view on the HST-ACS/WFC H$\alpha$ (+continuum) mosaic image (Mutchler et al.~2007).  Additionally, we show the resulting image (colour-inset) of the continuum-subtracted reconstructed H$\alpha$ image obtained with the IFU (the details are discussed below).

\subsection{Line fitting}

\begin{figure}
\begin{center}
\includegraphics[width=7cm]{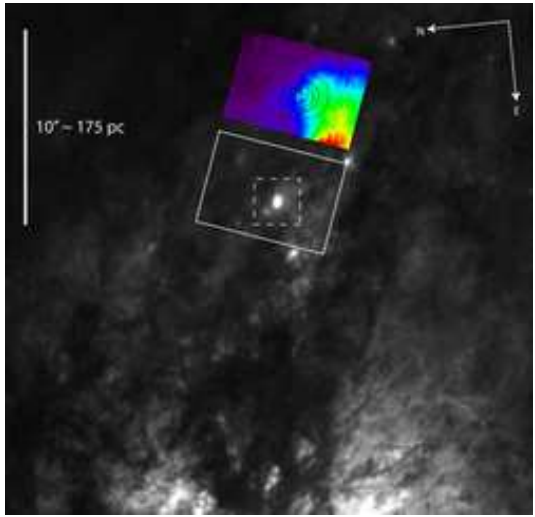}

\caption{An overlay of the GMOS-IFU field of view on the HST-ACS/WFC H$\alpha$ (+continuum) mosaic image.  The colour inset shows the reconstructed field of view of the H$\alpha$ intensity derived by fitting each spectrum in the datacube (see text for details).  The dashed square marks the region shown in Fig.~\ref{fig:colour-maps}.}

\label{fig:ifu-pointing}
\end{center}
\end{figure}

\begin{figure}
\begin{center}
\includegraphics[width=8cm]{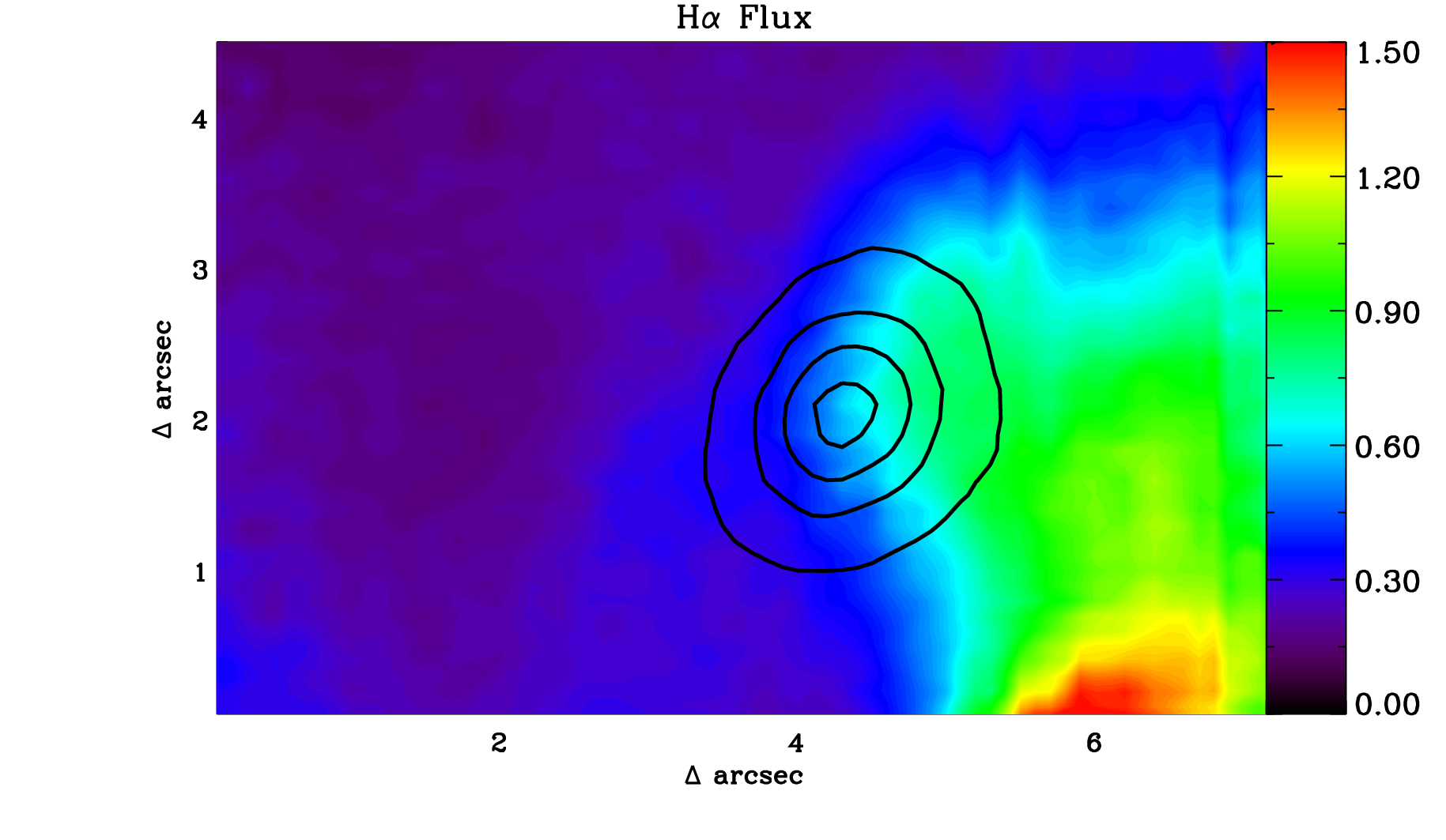}
\includegraphics[width=8cm]{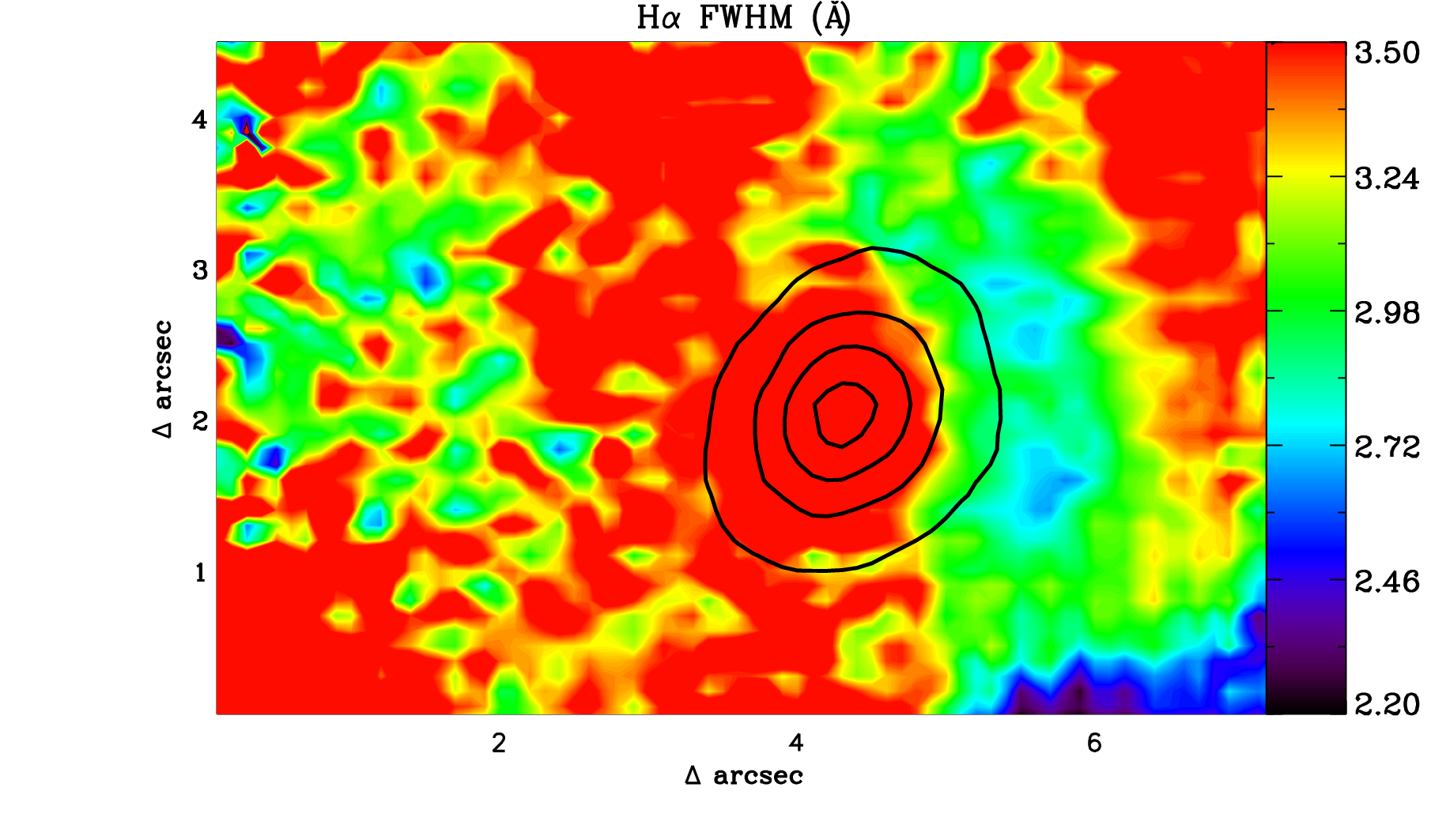}
\includegraphics[width=8cm]{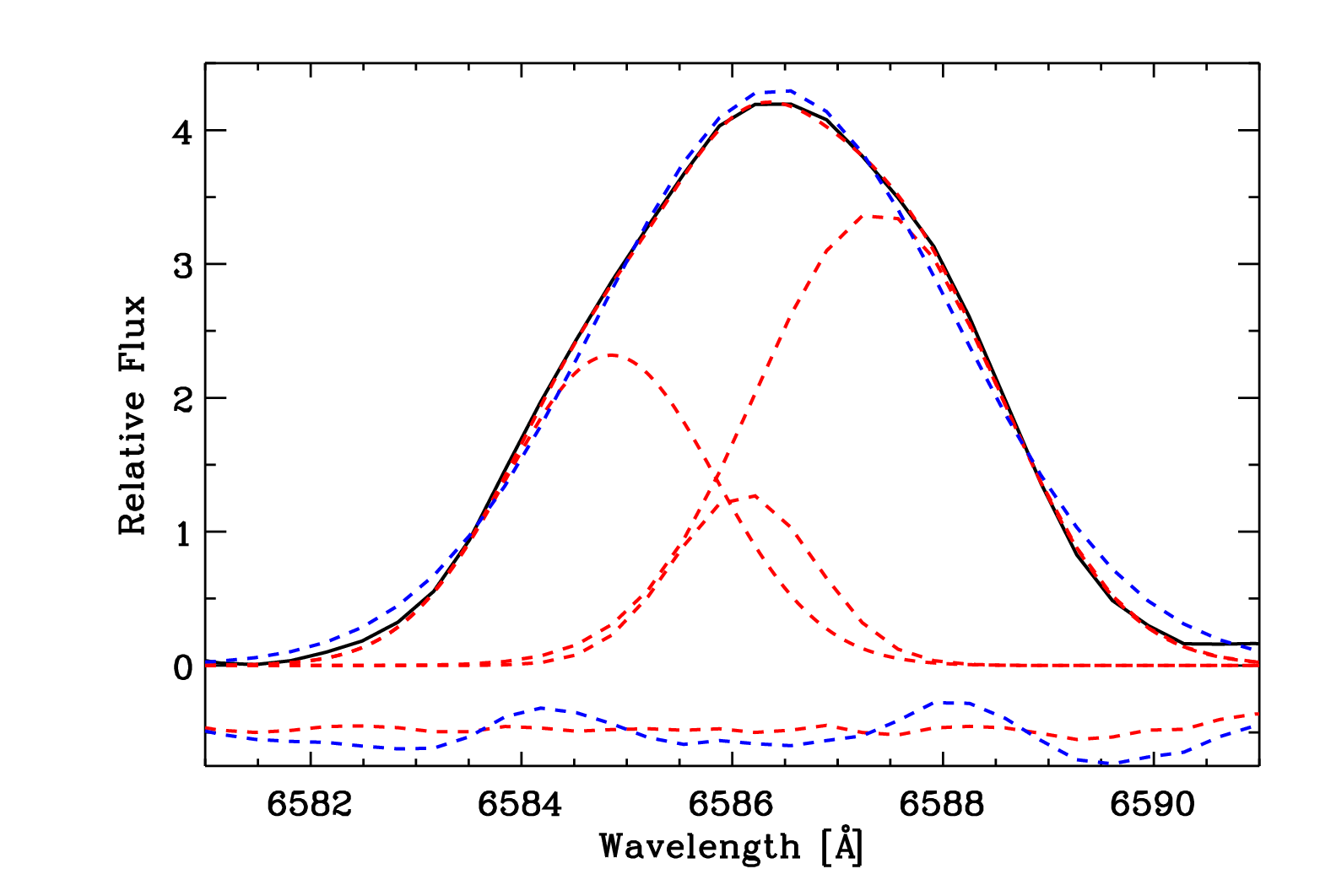}

\caption{{\bf Top:} The reconstructed map H$\alpha$ flux (continuum subtracted) from the GMOS IFU.  The position of M82F can be seen as contours which denoted 30, 50, 70, \& 90\% of the maximum continuum flux level.  {\bf Middle:} The same as the top panel, except now the  FWHM of the H$\alpha$ emission is shown (uncorrected from the instrumental resolution). {\bf Bottom:}  The line profile of [N~{\sc ii}] from the sum of all spatial pixels in a 1" by 1" box.  The observed profile is shown as a solid (black) line, while a single gaussian fit to the profile is shown as a (blue) dashed-dotted line.  Additionally, we show result of fitting three gaussians to the profile (each individual fit is shown as a red dashed line), and the sum of the three profiles is shown as a dashed line.  The difference between the observed spectrum (shifted down by 0.5 for clarity) and the fits are shown (using the same colour-coding as the fit).}

\label{fig:reconstructed}
\end{center}
\end{figure}

In order to see where the emission lines are strongest and to remove the continuum, we have fit single Gaussians to the H$\alpha$, [N~{\sc ii}], \& [S~{\sc ii}]  lines after subtraction of a linear fit to the continuum, using continuum bands just to the blue and red of the emission lines.  The reconstructed H$\alpha$ flux map is shown in the top panel of Fig.~\ref{fig:reconstructed}.  We see that the majority of the emission in the region is coming from an H{\sc ii} region just off the field of view of our IFU field.  A likely candidate star cluster is seen in Fig.~\ref{fig:ifu-pointing}.  M82F is located at the edge of the H{\sc ii} region and does not appear to be physically associated with the emission.

The reconstructed IFU image of the FWHM of the H$\alpha$ line (middle panel of Fig.~\ref{fig:reconstructed}, where we have not corrected for the instrumental resolution) shows some  unexpected features.  The line width varies across the field, with the H{\sc ii} region having a relatively narrow FWHM, and a (vertical) lane of broad emission which passes through the position of the cluster.  We note that this general behavior is also seen in the [N~{\sc ii}] and [S~{\sc ii}] lines.
 
In order to search for the cause of the broad lines, we have extracted a spectrum of all spatial pixels (spaxels) centred on the cluster in a 1" by 1" arcsecond box.  The resulting [N~{\sc ii}] line profile is shown in the bottom panel of Fig.~\ref{fig:reconstructed}, where structure in the profile suggests that multiple components are present.  This is expected from the high-resolution spectrum of SG01 who found three components in the H$\alpha$ emission profile of M82F.  To test whether three components are indeed present we have fit a single gaussian to each line profile along with a three component fit.  This is shown for [N~{\sc ii}] in the bottom panel of Fig.~\ref{fig:reconstructed}, where a single gaussian is shown as a dash-dotted line (blue) as well as a three gaussian fit along with their sum (red dashed lines).  The difference between the observed spectra and the fits are also shown. 

 In order to determine the robustness of the three component fit, we have also fit the H$\alpha$ and [S~{\sc ii}]\,$\lambda6716$~ in the same way.  We find similar velocities for each of the three components in each of the lines, namely 37 ($\pm27$), 117 ($\pm6$), and 189 ($\pm20$) km/s, where the values and errors represent the mean and standard deviation for each of the three lines.  The lowest velocity component agrees well with that derived from [O~{\sc iii}]\,$\lambda5007$~ in \S~\ref{sec:velocity}, while all three components are consistent with those found by SG01, hence confirming multiple-components are present.  However, it appears that none of the components are consistent with the velocity of the cluster.

The general structure of the FWHM reconstructed image is consistent with the extinction map derived in \S~\ref{sec:extinction} and seen in Fig.~\ref{fig:ext-map}.  Where the extinction is high, i.e. the lane of across M82F, we see multiple components to the emission lines (i.e. larger FWHM) and where the extinction is relatively low we see only a single component.  Thus, we are left to conclude that multiple layers of gas/dust are between us and M82F, and that the clouds which are causing the extinction in the vicinity of M82F are also emitting flux in the H$\alpha$, [N~{\sc ii}] and [S~{\sc ii}] lines, i.e. are at least partially ionised.

\section{Conclusions}\label{conclusions}

Using multi-band high-resolution images of the enigmatic cluster M82F
we present the following conclusions:

\begin{enumerate}

\item The integrated colours of M82F suggest an extinction of $>2.5$
  mag in $A_V$ and support an age of 60~Myr, consistent with the
  values measured by Smith \& Gallagher~(2001).
\item There is considerable sub-structure within the body of M82F as
  seen in the colour maps.  This substructure is also often seen in the
  observed contour maps.
\item By looking at the colour of each pixel in $(UV-B)$ vs. $(V-I)$
  space, we see that the pixels scatter parallel to the extinction
  vector, which supports the notion that the substructure is caused by
  differential extinction.
\item If one interprets this substructure as due to extinction (as
  opposed to age or population differences) then one can use the
  colour maps to create an extinction map of the cluster.  This map shows
  the existence of lanes of extinction through the cluster and also a
  wall of high extinction on the north-west side of the
  cluster. Differences in the amount of extinction are larger than 1.5~mag
  in A$_{V}$ in the region occupied by the cluster.  
\item Using this extinction map, it is possible to correct, on an
  individual pixel basis, the observed flux in order to derive the intrinsic flux.
  Doing this we see that the PA of the cluster can change by $\sim30$
  degrees (in the B-band within 20 pixels from the cluster centre).
\item Measuring the effective radius of M82F on the corrected image
  yields an increase of 15\% when using King profiles and $20-30\%$ when
  using EFF profiles. 
\item Adopting the EFF profile fits, we do not find any signature of a colour-size relationship that would indicate major mass segregation.  Mass segregation may, however, still be present, but simply not observable in this way.
 \item The PA and ellipticity of the cluster are much more stable in
 the intrinsic images than the observed images, showing the benefit of
 the extinction corrected images.

\end{enumerate}

Using the extinction corrected images of M82F, and fitting EFF
profiles to the cluster, we conclude that
mass-segregation is not observable in this cluster in the broad-band
images.  However, the increased size estimates of this cluster are not
enough to bring M82F into agreement with SSP models in the L/M$_{\rm
  dyn}$ diagram, leaving the origin of this discrepancy still
unknown.  It is possible that the stellar IMF within the cluster is
extremely top-heavy or that the large and spatially variable
extinction between the observer and the cluster causes complications
in the velocity dispersion measurements.  Another possibility is that
mass segregation, although not found in the broad-band data, causes an
under-estimation of the velocity dispersion, hence increasing the
L/M$_{\rm dyn}$ ratio.  Thus it remains to be seen if 
other clusters, with low extinction estimates, also fall in the region
of the L/M$_{\rm dyn}$ vs. age plane excluded by simple theory
(i.e. combining SSP models and the effects of rapid gas removal). 

From deep optical spectroscopy we have been able to derive the age of
M82F using the age dependent ratio of the depth of the
HeI$\lambda4471$ and MgII$\lambda4481$ lines.  The intrinsic benefits of
using this ratio to age date young star clusters are that it is largely
unaffected by reddening (due to the proximity of the lines in
wavelength), largely independent of metallicity, and very sensitive to
age in the range of $\sim40-400$~Myr. Using this technique we find an
age of $50-70$~Myr, assuming solar metallicity or $10-15$~Myr older if
one assumes $0.4~Z_{\odot}$.  Younger ages (i.e. $<20$~Myr) can
be ruled out by the widths of the Balmer lines.

  Finally, we have also presented Integral Field observations of M82F and its surrounding region.  We found that the majority of line emission seen in the region is not due to the cluster itself, but rather to a nearby H{\sc II} region.  We have found that the width of the emission lines corresponds spatially with the extinction map derived using optical colours, from which we conclude that the gas/dust clouds responsible for the extinction are also emitting in the H$\alpha$, [N~{\sc ii}], and [S~{\sc ii}] lines.  There is some evidence of weak additional emission lines associated with M82F, but we could not conclusively derive their origin.

\section*{Acknowledgments}

We would like to thank Anil Seth for discussions on the colour maps.  Barbara Ercolano is thanked for a critical reading of the manuscript.   NB gratefully acknowledges the hospitality of the Harvard-Smithsonian Center for Astrophysics where a large part of this research was carried out.
This paper is based on observations with the NASA/ESA {\it Hubble 
Space Telescope}\/ which is operated by the Association of Universities
for Research in Astronomy, Inc. under NASA contract NAS5-26555.  Based on observations obtained at the Gemini Observatory, which is operated by the Association of Universities for Research in Astronomy, Inc., under a cooperative agreement
with the NSF on behalf of the Gemini partnership.

\bsp
\label{lastpage}
\end{document}